\documentclass[aps,twocolumn,prb,superscriptaddress,showpacs,preprintnumbers,amsmath,amssymb]{revtex4}

\usepackage{graphicx}
\usepackage{color}
\usepackage{amsbsy}

\begin{document}

\title{Phase separation and second-order phase transition in the phenomenological model for Coulomb frustrated 2D system}

\author{R.~F.~Mamin}
\affiliation{Zavoisky Physical-Technical Institute, Russian
Academy of Sciences, 420029 Kazan, Russia}

\author{T.~S.~Shaposhnikova}
\affiliation{Zavoisky Physical-Technical Institute, Russian
Academy of Sciences, 420029 Kazan, Russia}

\author{V.~V.~Kabanov}
\affiliation{Department for Complex Matter, Jozef Stefan Institute, 1000 Ljubljana, Slovenia}
\affiliation{Zavoisky Physical-Technical Institute, Russian
Academy of Sciences, 420029 Kazan, Russia}

\date{\today}

\begin{abstract}
We have considered the model of the phase transition of the second order for the Coulomb frustrated 2D charged system. The coupling of the order parameter with the charge was considered as the local temperature. We have found that in such system, an appearance of the phase-separated state is possible. By numerical simulation, we have obtained different types ("stripes", "rings", "snakes") of phase-separated states and determined the parameter ranges for these states. Thus the system undergoes a series of phase transitions when the temperature decreases. First, the system moves from the homogeneous state with a zero order parameter to the phase-separated state with two phases in one of which the order parameter is zero and, in the other, it is nonzero ($\tau>0$). Then a first-order transition occurs to another phase-separated state, in which both phases have different and nonzero values of the order parameter (for $\tau<0$). And only a further decrease of temperature leads to a transition to a homogeneous ordered state.
\end{abstract}



\maketitle

\section{Introduction}
The problem of phase separation attracts considerable attention, because the variety of different phase states and the coexistence of several phases are observed in many materials currently being studied \cite{jin,Kugel,Kugel06,Dagotto01,Nagaev96,Tranquada04,Shen05,Blackburn13,Yamakawa15,Tranquada95}. This includes a class of manganites with colossal magnetoresistance \cite{jin,Kugel,Kugel06,Dagotto01,Nagaev96} in which there is a phase separation with charge inhomogeneity ("droplets", "bubbles", etc), as well as cuprate high-temperature superconductors \cite{Tranquada04,Shen05,Blackburn13,Tranquada95}, in which a pseudo gap state for $T>T_c$ and static and dynamic charge density waves (CDW) are observed.
The phenomenon of the phase separation is accompanied by the charge inhomogeneity, which is confirmed by various experimental observations. The charge inhomogeneity was observed by methods of scanning tunneling microscopy \cite{Neto14}, photoelectron spectroscopy with angular resolution (ARPES) \cite{Shen05},  X-ray and neutron diffraction \cite{Tranquada04,Tranquada95}.
For these materials, there is a certain range of temperature and doping level, in which the coexistence of phases is in the ground energy state. The spatial size of the single-phase regions is determined by the energy balance between the Coulomb interaction, which is important in the presence of an overcharge created by the doping, and the energy gain that appears when a more ordered phase occurs \cite{Mamin09,Shenoy07,Shenoy09}.

There are many theoretical studies of the states with charge inhomogeneity in which states with "droplets" and "stripes" has been obtained (see, for example, \cite{Kabanov09,Miranda08,Castro08,Castro08r,Castro09,Castro09a,Ortix07,Jamei,Muratov}). Usually in these papers it is considered the Coulomb frustrated first order phase transition where  the scalar order parameter is either coupled linearly with the charge density\cite{Castro08r,Ortix07,Jamei}, or the order parameter is proportional to the charge density\cite{Castro08,Castro08r,Castro09a,Muratov}. It is shown that these models are unstable with respect to phase separation. The phase-separated state represents the charged regions of different phases with different values of the order parameter.  Note that in the case of the second order phase transition this type of coupling of the order parameter to the charge density is forbidden. In the case of the second order phase transition the order parameter is not a scalar. Here we discuss the case of the Coulomb frustrated $second$ $order$ phase transition where we consider the lowest possible coupling of the charge density with the square of the  order parameter.
 Within this model we discuss a possibility of the existence of a phase-separated state with charge inhomogeneities near $T_c$, where in the matrix of the "high-temperature" phase with the order parameter equal to $\eta_1$ ($\eta_1 \neq 0$), exist inclusions of the "low-temperature" phase  with the order parameter equal to $\eta_2$ ($\eta_2>\eta_1$).  Moreover, with the change of temperature, several phase transitions can be observed.

In this paper we apply a phenomenological approach based on the Ginzburg-Landau theory to
describe the static phase separation of a 2D system in the vicinity of the second-order phase transition, taking into account the presence of the Coulomb interaction, associated with the overcharging effects due to doping. Because the types of materials stated above are quasi two dimensional (CuO planes in the cuprates and MnO  planes in a number of manganites), the 2D description represents a reasonable approximation. We define the range of parameters (related to the temperature and the doping), for which the phase separation is energetically favorable. We also calculate the region of the phase diagram in which two inhomogeneous phases coexist.

\section{Theoretical Model}
Let us consider the 2D system in the vicinity of the second order phase transition.
In de Gennes work \cite{deGennes60} the effects of the double exchange in compounds with mixed valency such as manganites $\rm (La_{1-x}Ca_x)({Mn_{1-x}}^{3+}{Mn_x}^{4+})O_3$ were studied. It was shown \cite{deGennes60} that the motion of "extra" holes or "extra" electrons in antiferromagnet is lowering the energy of the system.
Also it was shown that the Curie temperature depends on the doping $\rm x$.
Following de Gennes we begin with the Hamiltonian, where we add the term with the Coulomb interaction. For "layer" antiferromagnet the Hamiltonian may be written as following:

\begin{equation}
H=-\sum_{ij} J_{ij}{\bf S}_i {\bf S}_j
-\sum_{ij\sigma} t_{ij} a_{i\sigma}^+ a_{j\sigma} - J_H \sum_{i} {\bf S}_i {\bf s}_i
+
H_{Coul}
\label{EH1}
\end{equation}
Here the first term describe the exchange interaction of the $\rm Mn$ ions. ${\bf S}_i$ is the spin operator of the ionic spin on the  site ($i$). $J_{ij}$ is the exchange interaction, $J_{ij}$ connects only neighboring $i$ and $j$ magnetic sites.
The second and the third terms describe the double exchange interaction \cite{Izyumov01}. The second term in Eq. (\ref{EH1}) describes the hopping of an electron with the spin $\sigma$ along $ij$ lattice sites. $a_{i\sigma}^+$ ($a_{i\sigma}$) is the creation (annihilation) operator of electron on $i$ site, $t_{ij}$ is the
hopping integral. The third term of (\ref{EH1}) describes the Hund's coupling. Here ${\bf s}_i$ is  the spin operator of the conduction electron, which can be expressed in terms of the creation and annihilation operators for the electron
and the Pauli matrices \cite{Izyumov01}. The last term describes the Coulomb interaction.
Following de Gennes we assume that the spin ordering of the unperturbed system is of the "antiferromagnetic layer" type. Each ionic spin $S$ is coupled ferromagnetically to $z'$ neighboring spins in the same layer, and antiferromagnetically to $z$ spins in the adjacent layers. The exchange integrals are called $J'$($>0$) and $J$($<0$). The Zener carriers \cite{Zener59} are allowed to hop both in the layer (with transfer integrals $t'$) and also from one layer to the other (with transfer integrals $t$). The number of magnetic ions per unit volume is called $N$, and the number of Zener carriers $Nx$.
The model of double exchange is the exchange model under strong coupling conditions $J_H>>zt$ and $J_H>>z't'$.

In the limit of finite temperature and at low values of the relative sublattice magnetization, a phenomenological expression for the free energy was derived.
Then in the limit $J_H \rightarrow \infty$ the density of the thermodynamic potential of the system $\phi(\eta,\rho)$ ($\Phi=\int \phi(\eta,\rho) d\mathbf{r}$), which describes the order parameter $\eta$,  can be written in the form

\begin{equation}
\phi(\eta,\rho)=\phi_0+\phi_{\eta}+\phi_{int}+\phi_{Coul}\ ,
\label{E1}
\end{equation}
where for a phase transition of the second order

\begin{equation}
\phi_{\eta}=\frac{\alpha}{2}{\eta}^2+\frac{\beta}{4}{\eta}^4+
\frac{\delta}{6}{\eta}^6+
\frac{\zeta}{8}{\eta}^8+
\frac{D}{2}(\nabla \eta)^2.
\label{E2}
\end{equation}
The order parameter $\eta$ describes the relative magnetization of each sublattice \cite{deGennes60}.
Here $\alpha={\alpha}^{\prime}(T-T_c)$, $T_c$ is the phase transition temperature without doping,
${\alpha}^{\prime}\sim 1/C$, $C$ is the Curie constant, $\beta>0$.
$\phi_{\eta}$  includes a second-order term from $\eta$, a positive fourth-order, a positive sixth-order, a positive eighth-order term and a gradient term. Here:

\begin{equation}
\alpha=2N (\frac{3}{2} k_{B} T - S^2 (z'J'+ z J))
\label{EH8},
\end{equation}

\begin{equation}
\beta=4N (\frac{9}{20} k_{B} T + \frac{6}{175} x(z't'+ z t)),
\label{EH10}
\end{equation}

\begin{equation}
\delta=6N (0.325 k_{B} T + 0.27 x(z't'+ z t)),
\label{EH11}
\end{equation}

\begin{equation}
\zeta=8N (0.06 k_{B} T + 2.21 x(z't'+ z t)),
\label{EH12}
\end{equation}
$k_{B}$ is the Boltzmann's constant.
$\phi_{int}$ describes the interaction of the order parameter $\eta$ with the local
charge density ${\rho}$
\begin{equation}
\phi_{int}=-\frac{\sigma_1}{2}{\eta}^2 \rho.
\label{E3}
\end{equation}
The expression is obtained from the double exchange interaction terms (see Eq.(\ref{EH1})) averaged over the temperature. The interaction is written here as the local temperature,
$\sigma_1$ is the interaction constant.

Main physical properties of the system are determined by the parameter $\sigma_1$, that is defined as
\begin{equation}
\bar{\rho} \sigma_1=\frac{4N}{5} x(z't'+ z t).
\label{EH9}
\end{equation}

\begin{equation}
\phi_{Coul}=\frac{\gamma}{2}\int \frac{(\rho(\mathbf{r})-\bar{\rho})(\rho(\mathbf{r}^\prime)-\bar{\rho})}
{\mid \mathbf{r}-\mathbf{r}^\prime\mid} d \mathbf{r}^\prime
\label{E4}
\end{equation}
is the energy density of the Coulomb interaction, the constant $\gamma$ is determined by the dielectric constant. In the absence of terms of $\phi_{int}$ and $\phi_{Coul}$
a  phase transition of the second order is observed at $\alpha=0$. For $\alpha<0$ there exists an equilibrium value of the order parameter $\eta \ne 0$. For $\alpha > 0$, the equilibrium value of $\eta = 0$, then there is no order, which is determined by the parameter $\eta$. In expressions (\ref{EH9}) and (\ref{E4}) $\bar{\rho}$  is the average 2D surface density of charge
\begin{equation}
\bar{\rho}=\frac{1}{S} \int_{S} \rho d\mathbf{r},
\label{}
\end{equation}
where $\mathbf{r}$ is 2D-vector.

The total free energy $\Phi$ should be minimized with respect to $\eta(\mathbf{r})$ and $\rho$. The minimization of $\Phi$ with respect to $\rho$ gives

\begin{equation}
-\frac{\sigma_1}{2}{\nabla}^2_{3D} {\eta}^2
=4\pi \gamma(\rho(\mathbf{r})-\bar{\rho})\delta(z)d .
\label{E6}
\end{equation}
Here  thickness  of 2D-layer  $d$  is introduced to preserve dimensionality.
$\delta(z)$ is the Dirac delta-function. Substituting  (\ref{E6}) in (\ref{E1}), we obtain

\begin{eqnarray}
\phi & = & {\phi}_0 +\frac{\alpha}{2}{\eta}^2+\frac{\beta}{4}{\eta}^4
+\frac{\delta}{6}{\eta}^6+
\frac{\zeta}{8}{\eta}^8 +
\nonumber\\
& & +\frac{D}{2}(\nabla \eta)^2-\frac{\sigma_1}{2} {\eta}^2 \bar{\rho}-
\nonumber\\
& & -\frac{{\sigma_1}^2}{32 {\pi}^2 \gamma d^2}\int \frac{\nabla_{2D} {\eta}^2(\mathbf{r})\nabla_{2D}
{\eta}^2(\mathbf{r}^\prime)}{\mid \mathbf{r}-\mathbf{r}^\prime\mid} d \mathbf{r}^\prime ,
\label{E7}
\end{eqnarray}
where $\mathbf{r}$ and $\mathbf{r}^\prime$ are 2D vectors. The last two terms in this expression are negative. The term $\frac{\sigma_1}{2} {\eta}^2 \bar{\rho}$ renormalizes
the critical temperature of the phase transition. The critical temperature now depends on the average charge density. The coefficient in front of ${\eta}^2$ is changed from
$\alpha$ to $\tilde{\alpha}$.

\begin{equation}
\tilde{\alpha}=\alpha -\sigma_1 \bar{\rho}
\end{equation}

Note that the presence of last nonlocal term in expression (\ref{E7}) leads to the instability of the homogeneous state.

Let us introduce new dimensionless values  $\Lambda$ and $\boldsymbol{\xi}$ as
$\Lambda=\eta/\eta_0$ and $\xi_x=x/a$, $\xi_y=y/a$, where ${\eta_0}^4=\beta /\zeta$
 and $a=\sqrt{\frac{D {\zeta}^{1/2}}{2{\beta}^{3/2}}} \chi$. Here $\chi$ is a constant.
 We choose the value of the constant in the interval from $3$ to $20$. This constant allows us to change the size of an area in which the spatial distribution of the order parameter $\eta(r)$ is calculated. Then the expression (\ref{E7}) has the form

\begin{eqnarray}
\phi & = & U_0 \biggl( \tau {\Lambda}^2 + \frac{{\Lambda}^4}{2} + \tilde{\delta} \frac{{\Lambda}^6}{3}+
\frac{{\Lambda}^8}{4}+
\frac{2}{{\chi}^2} (\nabla \Lambda)^2 -
\nonumber\\
& & - \frac{A}{\chi} \int \frac{\nabla_{2D} {\Lambda}^2(\boldsymbol{\xi})\nabla_{2D}
{\Lambda}^2(\boldsymbol{\xi}^\prime)}{\mid \boldsymbol{\xi}-\boldsymbol{\xi}^\prime\mid} d \boldsymbol{\xi}^\prime \biggr)
\label{E9}.
\end{eqnarray}
Here parameters $U_0$, $\tau$, $\chi$, $A$ and $\tilde{\delta}$ are defined as

\begin{equation}
U_0 = \frac{\beta}{2} {\eta_0}^4 = \frac{\beta^2}{2\zeta},
\label{E11}
\end{equation}

\begin{equation}
\tau=\frac{\tilde{\alpha}}{\beta {\eta_0}^2}=\sqrt{\frac{\zeta}{\beta^3}}\bigl( {\alpha}^{\prime}(T-T_c)-\sigma_1 \bar{\rho}
\bigr),
\label{E12}
\end{equation}

\begin{equation}
\chi= a \eta_0 \sqrt{\frac{2\beta}{D}},
\label{E13}
\end{equation}

\begin{equation}
A=\frac{{\sigma_1}^2}{8 \gamma d^2 {\pi}^2 \sqrt{2D} \sqrt[4]{\beta \zeta} },
\label{E14}
\end{equation}

\begin{equation}
\tilde{\delta}=\frac{\delta}{\beta} {\eta_0}^2=\frac{2 \delta}{\sqrt{\beta \zeta}}.
\label{E15}
\end{equation}

\section{Results}
In order to find the minimum of $\Phi=\int \phi d\mathbf{r}$ (\ref{E9}) the method of conjugate gradients (CGM) was used. We have introduced $\mathcal{N} \times \mathcal{N}$ ($\mathcal{N}=128$ or $\mathcal{N}=512$) discrete points on a square with side $a$. We have applied the periodic boundary conditions. In the numerical calculations three parameters $A$, $\tau$ and $\chi$ were used.

\begin{figure}[t]
\centering{\includegraphics[width=0.8\columnwidth,angle=0,clip]{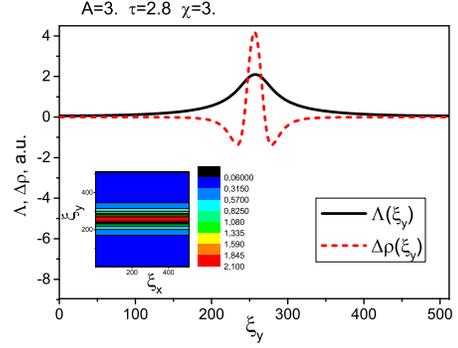}}
\caption{\label{fig1}
The distribution of the order parameter $\Lambda (\xi_x=const,\xi_y)$(solid curve) and
$\Delta \rho=\rho(\xi_x=const,\xi_y)-\bar{\rho}$ (dashed curve) for $\chi=3$, $A =3$ and $\tau=2.8$ in inhomogeneous state.
The inset shows the distribution of the order parameter in 2D. The figure represents the results of numerical calculations for $\mathcal{N}=512$.}
%Color contours from highest to lowest values of the order parameters
%follow the red-orange-yellow-green-blue degradation (color contours are kept in the next figures).}
\end{figure}

We have studied the dependence of the free energy from the parameters $A$ and $\tau$ with the fixed value of $\chi$. The inset in Fig.\ref{fig1} shows the spatial distribution of the order parameter $\Lambda$($\xi_x, \xi_y$) for the parameters
$A=2.8$, $\tau=3$, $\chi=3$ and $\mathcal{N}=512$. The free energy of this state is negative ($\Phi<0$). The inset shows that at the given values of the parameters a phase separation takes place. It means that spatially inhomogeneous distribution of the order parameter represent a global minimum of the free energy. This state is energetically more favorable than the homogeneous state (the energy of homogeneous state with
$\Lambda(\mathbf{r})=0$ is $\Phi=0$). These non-uniform states are formed because of the  charge redistribution \cite{Mamin09}.

Fig.\ref{fig1} shows the distribution of the order parameter $\Lambda$($\xi_x$, $\xi_y$)
and the incremental charge $\Delta \rho=\rho(\xi_x,\xi_y)-\bar{\rho}$ along the line perpendicular to the strip (along y axis). As it follows from this figure,
in the region of inhomogeneous distribution of the order parameter $\Lambda$($\xi_x$=const, $\xi_y$) there exists a triple extra charged layer. The total charge of this layer is zero with high precision, $\Delta \rho>0$  in the center of a stripe, and $\Delta \rho<0$ on each side (dashed curve).

For the fixed values of the parameter $A=3$  the inhomogeneous distribution of the order parameter exists in the range $\tau_2 \le \tau \le \tau_3$ ($\tau_2=-27$ and $\tau_3=4.2$ for $A=3$). And the free energy is less than zero for  $\tau \le \tau_1$ ($\tau_1=3.3$ for $A=3$).

According to Eq.(\ref{E12}) $\tau$ is a linear function of $T-T_c$ and is changed with $\bar{\rho}$, where $T_c$ is the transition temperature in the absence of interaction (i.e. at $\Phi_{int}=0$). $\bar{\rho}$  is the average charge, $\bar{\rho}$ is proportional to the level of doping. The parameter $A$ Eq.(\ref{E14}) depends on the coupling parameter
$\sigma_1$ and the strength of the Coulomb interaction. With the increase of the Coulomb interaction parameter $A$ decreases. As a result the region of $\tau $, where the phase separation is observed, is shrinking.

\begin{figure}[t]
\centering{\includegraphics[width=2.
\columnwidth,angle=0,clip]{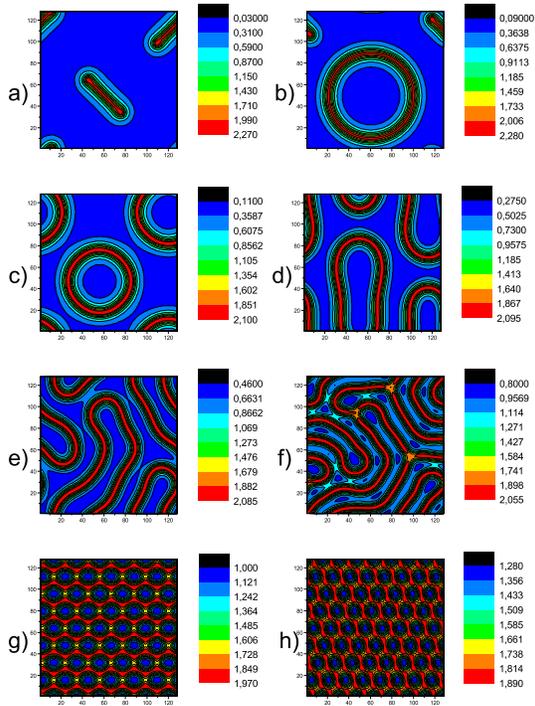}}
\caption{\label{fig7}Inhomogeneous states are shown for $A=3$, $\chi=5$ and $\tau=3.2(a), 2(b), 1(c), -1(d)$,
$-5(e), -15(f)$, $-20(g), -25(h)$, respectively. The phase-separated state is stable at $3.2 \leq \tau \leq -27$. The parameter $\tau$ decreases from a) to h), corresponding to the decreasing of temperature $T$. All figures represent the results of numerical calculations for $\mathcal{N}=128$. The order parameter changes from $\Lambda_{min}=0$ to $\Lambda_{max}=2.2$ in Figures a) and b). The difference between $\Lambda_{max}$ and $\Lambda_{min}$ decreases from Figure c) to Figure h).}
\end{figure}

In Fig.\ref{fig7} a change of a form of inhomogeneous states is shown for $A=3$ and with the reduction of $\tau$ from $3.2$ to $-25$.
The free energy of these inhomogeneous states is negative and smaller than the energy of an homogeneous state.

The landscape of the phase separation changes with the change of $\tau$ as shown on the Fig.\ref{fig7}. For $\tau>0$ the phase separation is observed in the form of stripes or rings. The stripe with $\Lambda>0$  appears on the background with zero order parameter $\Lambda=0$ . The stripes may be straight or may have more complex closed form. With increasing of $\tau$, the number of such stripes is reduced, and the rings are compressed. Note that the value of the order parameter in the center of the stripes is not changed (see Fig.\ref{fig7} c,b,a).
When the value of $\tau$ becomes negative and with the further reduction of $\tau $, the loop's form is changing. They are  bent more strongly, and the value of the order parameter in the "background" becomes different from zero.
With further decreasing of $\tau$ the phase separation becomes more shallow. The difference  between $\Lambda$  inside and outside of the "stripe" is decreasing to zero at $\tau = \tau_2$ and the transition to the homogeneous state with $\Lambda$=const occurs
(see Fig.\ref{fig7}d-h).

\begin{figure}[t]
\centering{\includegraphics[width=2.
\columnwidth,angle=0,clip]{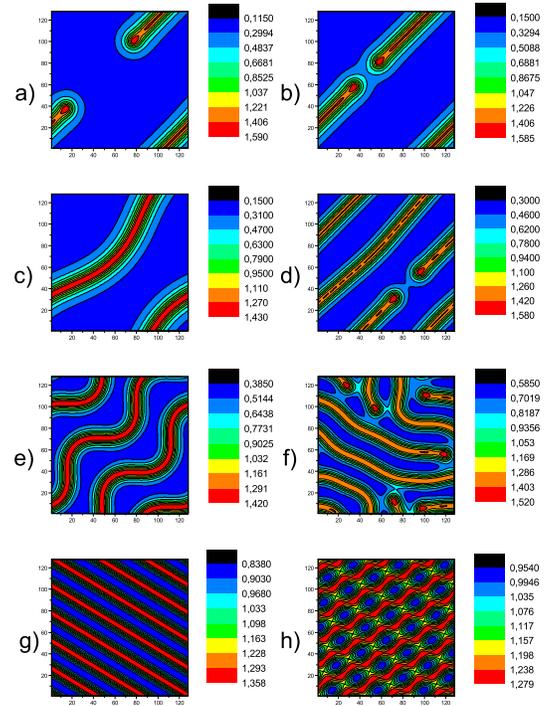}}
\caption{\label{fig8}Inhomogeneous states are shown for $A=2.2$, $\chi=10$ and $\tau=0.01(a), -0.1(b), -0.2(c), -0.5(d)$,
$-1(e), -2(f)$, $-3(g), -3.7(h)$, respectively. The inhomogeneous state exists a more narrow interval of $\tau$ ($0.01\leq \tau \leq - 0.37$) for $A = 2.2$  than for $A = 3$.
The decrease of $A$ (increase of Coulomb interaction) leads to the decrease of the interval of $\tau$ where the phase separation is observed. All figures represent the results of numerical calculations for $\mathcal{N}=128$.}
\end{figure}

\begin{figure}[t]
\centering{\includegraphics[width=0.9\columnwidth,angle=0,clip]{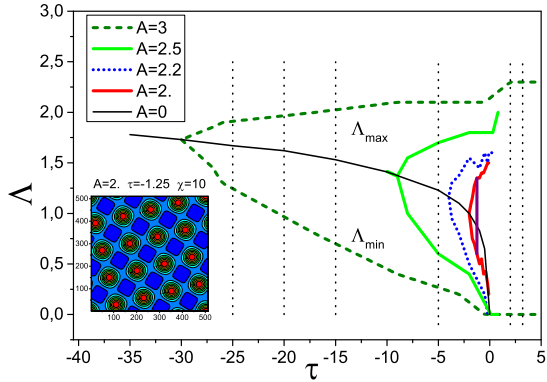}}
\caption{\label{fig9}The maximum $\Lambda_{max}$ and minimum $\Lambda_{min}$ values of the order parameter in the inhomogeneous states as a function of $\tau$ for different values of $A$.
The inset shows the distribution of the order parameter $\Lambda$ in 2D for $A=2.$, $\tau=-1.25$ and $\xi=10.$ The figure represents the results of numerical calculations for $\mathcal{N}=512$.}
\end{figure}

In Fig.\ref{fig8} the change of a form of the inhomogeneous states is shown for $A=2.2$,
$\chi=10$ and with the reduction of $\tau$ from $0.01$ to $-3.7$. From figures \ref{fig7} and \ref{fig8} it is clearly seen that the main features of a phase separated state are similar. Note that the region of the existence of a phase separated state for $A=2.2$ is reduced in comparison with that for $A=3.$ (see Fig.\ref{fig9}).

Figure \ref{fig9} shows the dependence of the maximum value of the order parameter
$\Lambda_{max}$ and the minimum value $\Lambda_{min}$ as a function of $\tau$ for four different values of $ A =$ 2 , 2.2, 2.5, and 3 in the phase-separated state.
A smooth solid line shows the change in the order parameter for $A = 0$, i.e. for the case when there is no interaction between the order parameter and the charge
($\sigma_1 = 0$, see Eq. (\ref{E3})).
Fig.\ref{fig9} shows that the phase transition of the second order is observed at $A = 0$ and $\tau = 0$. The order parameter is zero for $\tau>0$, and the phase with a nonzero order parameter appears below $T_c$ $\tau<0$. The energy of this state becomes negative $\Phi_{hom} <0$ at $\tau<0$.

In our model there is an interaction of the order parameter and the charge
($ \sigma_1 \neq 0 $).
Therefore, the minimum in free energy $\Phi_{inhom} <0$ corresponds to an inhomogeneous phase-separated state with the order parameter varying from $\Lambda_{min}$ to $ \Lambda_{max} $ (see Fig.\ref{fig9}). Consider the changes of phases that occur when $\tau$ decreases for the case of $ A = 3$. An inhomogeneous phase-separated state appears as a jump (a phase transition of the first order) at $\tau = \tau_1$. The regions with
$\Lambda \neq 0$ grows on the background with zero order parameter $\Lambda = 0$.
$\Lambda_{max}$=2.2 in these regions.  The number of such regions increases when $\tau $ decreases from $\tau_1 $ to $0$. Note that the values of $ \Lambda_{max} = 2.2 $ and $ \Lambda_{min} = 0 $ do not change in this region of $\tau$ (see Fig.\ref{fig7}a,b,c).
At $\tau = 0$, the phase-separated state starts to change. $\Lambda_{max}$ begins to decrease, and $\Lambda_{min}$ begins to increase (see Fig.\ref{fig7}d-h).
With further decreasing of $\tau <0$, the difference between $\Lambda_{max}$ and
$\Lambda_{min}$ decreases, and  $\Lambda_{max} = \Lambda_{min} = \Lambda$ at $\tau =\tau_2$, therefore a phase transition of the second order from an inhomogeneous state to a homogeneous state is observed.

\begin{figure}[t]
\centering{\includegraphics[width=0.8\columnwidth,angle=0,clip]{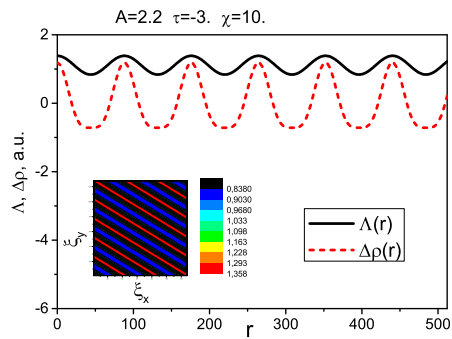}}
\caption{\label{fig33}
The distribution of the order parameter $\Lambda (r)$(solid curve) and
$\Delta \rho=\rho(r)-\bar{\rho}$(dashed curve) for $\chi=10$, $A =2.2$ and  negative value $\tau=-3$ in the inhomogeneous state.
The direction of $r$ is chosen perpendicular to the strips on the inset.
The inset shows the distribution of the order parameter in 2D for this set of parameters.
The figure represents the results of numerical calculations for $\mathcal{N}=128$.}
\end{figure}

Fig.\ref{fig33} shows the inhomogeneous distribution of the order parameter $\Lambda$($r$)
and the incremental charge $\Delta \rho=\rho(r)-\bar{\rho}$ along a line perpendicular to strips for negative value $\tau=-3$. The order parameter $\Lambda$ is varying from
$\Lambda_{min}=0.8$ to $\Lambda_{max}=1.3$.
In the region of inhomogeneous distribution of the order parameter $\Lambda$ there exists
inhomogeneous distribution of the incremental charge (dashed curve).

When $\chi$ changes from $3$ to $20$ (with $A=const$), the interval of $\tau$ where inhomogeneous states are formed does not change.

\begin{figure}[t]
\centering{\includegraphics[width=1.8\columnwidth,angle=0,clip]{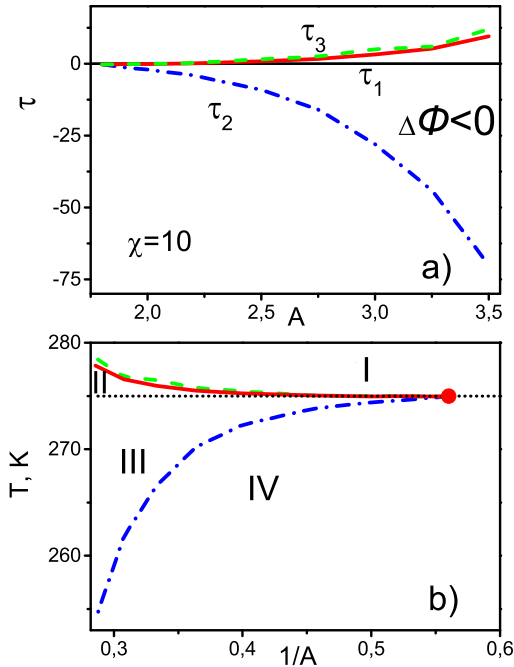}}
\caption{\label{fig3}a) The phase diagram of inhomogeneous states in axes
$A-\tau$ for $\chi=10$ for which $\Delta \Phi<0$. The energy of the inhomogeneous phase-separated state $\Phi_{inhom}$ is lower than the energy of the homogeneous state $\Phi_{hom}$ at $ \tau_2<\tau<\tau_1 $. $\Delta \Phi =\Phi _{inhom}-\Phi _{hom}$.
b)	The phase diagram plotted in the axes $T-1/A$, where $T$ is the temperature. The following parameters were used: $T_c + \sigma_1 \bar{\rho} = 275$K, $\frac{\tau}{\alpha'}\sqrt{\frac{\beta^3}{\zeta}}$ = 0.3K$^{-1}$. Region I is a homogeneous non-magnetic state with $\Lambda=0$. Region II is a phase-separated state with zero and nonzero order parameters. Region III is a phase-separated state with  nonzero order parameter. Region IV is the homogeneous magnetic state with $\Lambda\neq0$. The point at $A=1.8$ is the critical end point. For $A<1.8$ the phase separation is impossible.}
\end{figure}

In Fig.\ref{fig3}a $\tau=\tau_1$ and $\tau=\tau_2$ lines indicate the boundaries of the inhomogeneous states in axes
$A-\tau$ for $\chi=10$ for which $\Phi<0$.
The figure shows that with the increase of the parameter $A$ the regions of $\tau$ in which inhomogeneous distribution of the order parameter $\Lambda(\xi_x,\xi_y)$  was observed, is expanding.
In Fig.\ref{fig3}a the $\tau=\tau_3$  line  shows the boundary of the region of metastable inhomogeneous phases. For $\tau_1<\tau<\tau_3$
heterogeneous state corresponds to the local minimum of the free energy, but $\Phi>0$. This state is similar to "superheated liquid".

In Fig.\ref{fig3}b a phase diagram of inhomogeneous states is shown in the axes $T-1/A$. I and IV regions correspond to the homogeneous phases with zero and nonzero order parameters, respectively. II and III regions correspond to the inhomogeneous phases. $1 /A$ is proportional to the value of the Coulomb interaction $\gamma$ and inversely proportional to the square $\sigma_1$ (see Eq.(19)).  The phase separation is impossible below the critical end point at $A=1.8$ which is represented by the dot in the phase diagram Fig\ref{fig3}b. Indeed at a large value of  the Coulomb interaction and a small value of the double-exchange energy the Coulomb energy for any charge modulation becomes so large that it is always larger than energy gain due to ordering.  

%Increase of the interaction constant $\sigma_1$ leads to the increase of the parameter $A$. In accordance with Fig.\ref{fig3} that leads to the %increase of the temperature interval  where a phase separation is observed.
%There is a certain threshold value of the  interaction constant, below which the
%phase separation is absent (see Fig.\ref{fig3}). The level of doping influences the offset  of the temperature interval, in which  phase separation is %observed.
%he increase of the level of doping is (i.e. increase of $\bar{\rho}$) leads to decrease  of the transition temperature to the phase-separated state %and to the spatially uniform state with the nonzero order parameter.

\begin{figure}[t]
\centering{\includegraphics[width=0.9\columnwidth,angle=0,clip]{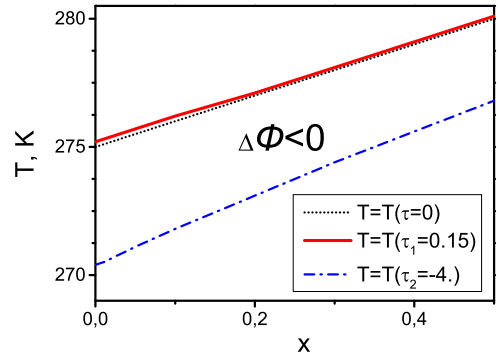}}
\caption{\label{fig10}Phase diagram in $x-T$ axes. The parameter $A$ is equal to $2.7$, and $\sigma_1=10$. The area between solid and dash dot lines is a phase-separated state, which corresponds to regions II and III in Fig. 6b. Dot line is the temperature of the phase transition in the absence of a phase-separated state.}
\end{figure}

Figure \ref{fig10} shows the phase diagram of the inhomogeneous state in $x-T$ axes for $A=2.7$ and $\sigma_1=10$. The decrease in $A$ leads to the decrease of the area where the phase separation is observed. 

As it was mentioned in the Introduction the phase separation is observed in manganites as well as in the cuprate high-temperature superconductors 
\cite{jin,Kugel,Kugel06,Dagotto01,Nagaev96,Tranquada04,Shen05,Blackburn13,Yamakawa15,Tranquada95}.
We discuss in this paper the inhomogeneous phases in manganites where a sequence of phase
transitions to inhomogeneous states is observed \cite{Kugel,Kugel06,Dagotto01}. Let us 
consider La$_{1-x}$Sr$_{x}$MnO$_3$ system. For strontium concentration $x=0.125$ the 
following sequence of the phase transition is observed.  First at $T=275$K the transition 
from homogeneous to inhomogeneous phase I is observed.  Then with lowering of the 
temperature the transition to inhomogeneous phase II takes place. And only then at 140K 
the system undergoes the transition to homogeneous state \cite{Dagotto01,Eremina,MaminNJP}. This sequence of the phase transitions is very similar 
to that discussed in our paper. In addition  similar inhomogeneous states may 
appear in the cuprates as well \cite{Fradkin12,PRL_002,Ortix06,Lorenzana03,DiCastro05}.

\section{Conclusion}
In conclusion we consider the theory of phase transition of the second order, where in addition to the standard expansion of the free energy in powers of the order parameter, it was introduced the Coulomb interaction and the interaction of a charge with the order parameter. The distribution of the order parameter and the charge distribution in 2D plane, that correspond to the minimum of free energy were found.
Numerical calculations were performed using the CGM method. Calculations showed that between the regions which are characterized by constant values of the order parameter, there is an area with inhomogeneous distribution of the order parameter and inhomogeneous distribution of the charge. This phase separation can exist in a form of one-dimensional stripes or in two dimensional  rings or "snakes".
A series of phase transitions have been found. With a decrease of temperature, first, the phase transition from the homogeneous state with zero order parameter to the phase-separated state with two phases with zero and nonzero order parameter ($\tau>0$) occurs. Then a first-order phase transition to another phase-separated state, in which both phases have different and nonzero values of order parameter (for $\tau<0$), is observed. And only with a further decrease of temperature the transition to a homogeneous ordered state takes place.
The regions in the parameter space "temperature-doping level", for which the phase separation or the coexisting of phases occur, have been defined.
We have tracked the change in the type of phase separation under the change of the temperature, the doping level of the material and  the alteration of the coupling constant.

\begin{acknowledgements}
The authors are grateful to A.V. Leontiev for numerous illuminating discussions.
One of us R.F.M. acknowledges the financial support from Slovenian Research Agency, Project BI-RU/16-18-021.
\end{acknowledgements}

%\end{thebibliography}
%\bibliography{stati_en}

\end{document}